\documentclass[11pt]{article}

\usepackage{fullpage,amsmath,amsfonts,latexsym}

\def\01{\{0,1\}}

\newcommand{\ket}[1]{|#1\rangle}

\newcommand{\EQP}{\mbox{\rm EQP}}
\newcommand{\NQP}{\mbox{\rm NQP}}
\newcommand{\ZQP}{\mbox{\rm ZQP}}
\newcommand{\BQP}{\mbox{\rm BQP}}
\newcommand{\ZPP}{\mbox{\rm ZPP}}
\newcommand{\AND}{\mbox{\rm AND}}
\newcommand{\OR}{\mbox{\rm OR}}

\newtheorem{theorem}{Theorem}
\newtheorem{lemma}{Lemma}

\newenvironment{proof}
{\noindent {\bf Proof. }}
{{\hfill $\Box$}\\
 \smallskip}

\begin{document}

\title{Quantum Zero-Error Algorithms Cannot be Composed\thanks{Partially 
funded by projects QAIP (IST--1999--11234) and RESQ (IST-2001-37559)
of the IST-FET programme of the EC.}}
\author{Harry Buhrman\\ 
CWI and U.~of Amsterdam\\{\tt buhrman@cwi.nl} 
\and Ronald de Wolf\\ 
CWI\\{\tt rdewolf@cwi.nl} 
}
\date{}
\maketitle

\begin{abstract}
We exhibit two black-box problems, both of which have an efficient 
quantum algorithm with zero-error, yet whose composition does not have
an efficient quantum algorithm with zero-error.
This shows that quantum zero-error algorithms cannot be composed.
In oracle terms, we give a relativized world where $\ZQP^{\rm ZQP}\neq \ZQP$,
while classically we always have $\ZPP^{\rm ZPP}=\ZPP$. \\[2mm]
Zero-error computation.
\end{abstract}

\section{Introduction}

We can define a ``zero-error'' algorithm of complexity $T$ in
two different but essentially equivalent ways: 
either as an algorithm that 
always outputs the correct value with {\em expected\/} 
complexity $T$ (expectation taken over the internal 
randomness of the algorithm), or as an algorithm
that outputs the correct value with probability at least $1/2$,
never outputs an incorrect value, 
and runs in {\em worst-case\/} complexity $T$.
Expectation is linear, so we can compose two classical 
algorithms that have an efficient expected complexity to
get another algorithm with efficient expected complexity.
If algorithm $A$ uses an expected number of $a$ applications 
of $B$ and an expected number of $a'$ other operations,
then using a subroutine for $B$ that has an expected number 
of $b$ operations gives $A$ an expected number 
of $a\cdot b+a'$ operations.
In terms of complexity classes, we have 
$$
\ZPP^{\rm ZPP}=\ZPP,
$$
where $\ZPP$ is the class of problems that can be solved by 
a polynomial-time classical zero-error algorithm.
This equality clearly relatives, i.e., it holds relative to
any oracle $A$. 

In this paper we show that this seemingly obvious composition fact 
does {\em not\/} hold in the quantum world. 
We exhibit black-box (query complexity) problems $g$ and $h$ 
that are both easy to quantum compute in the expected sense, 
yet whose composition $f=g(h,\ldots,h)$ requires 
a very large expected number of queries. 
In complexity terms, we exhibit an oracle $A$ where 
$$
\ZQP^{\rm ZQP^A}\neq \ZQP^A,
$$
where $\ZQP$ is the class of problems that can be solved by 
a polynomial-time quantum zero-error algorithm.
This result is somewhat surprising, because {\em exact\/}
quantum algorithms can easily be composed, and so can 
{\em bounded-error\/} quantum algorithms. Moreover, it is also
easy to use a quantum zero-error algorithm as a subroutine 
in a {\em classical\/} zero-error algorithm.
That is
$$
\EQP^{\rm EQP}=\EQP 
\mbox{ \ and \ }\BQP^{\rm BQP}=\BQP 
\mbox{ \ and \ }\ZPP^{\rm ZQP}=\ZQP,
$$
relativized as well as unrelativized.

\section{Preliminaries}

We assume familiarity with 
computational complexity theory~\cite{papadimitriou:cc} and
quantum computing~\cite{nielsen&chuang:qc}.
In this section we briefly introduce the ``modes of computation''
that we are considering. Let $f$ be some (possibly partial) 
Boolean function with set of inputs ${\cal X}={\cal X}_0\cup{\cal X}_1$,
where $f({\cal X}_0)=0$ and $f({\cal X}_1)=1$.
Let $P_b(x)$ be the probability that algorithm $A$ outputs 
bit $b$ on input $x$. We define four modes of computation:
\begin{enumerate}
\item $A$ is an {\em exact\/} algorithm for $f$ 
if $P_1(x)=1$ for all $x\in{\cal X}_1$ and 
$P_0(x)=1$ for all $x\in{\cal X}_0$
\item $A$ is a {\em zero-error\/} algorithm for $f$ 
if $P_1(x)\geq 1/2$ and $P_0(x)=0$ for all $x\in{\cal X}_1$
(assume there is a third possible output ``don't know''), and
$P_0(x)\geq 1/2$ and $P_1(x)=0$ for all $x\in{\cal X}_0$
\item $A$ is a {\em bounded-error\/} algorithm for $f$ 
if $P_1(x)\geq 2/3$ for all $x\in{\cal X}_1$, and
$P_0(x)\geq 2/3$ for all $x\in{\cal X}_0$
\item $A$ is a {\em nondeterministic\/} algorithm for $f$ 
if $P_1(x)>0$ for all $x\in{\cal X}_1$, and $P_1(x)=0$ for all $x\in{\cal X}_0$
\end{enumerate}
Note that an exact algorithm is a zero-error algorithm,
and a zero-error algorithm is a bounded-error algorithm
as well as a non-deterministic algorithm.

In the setting of {\em query complexity}, $f$ is an $N$-bit Boolean function,
so ${\cal X}_0\cup{\cal X}_1\subseteq\01^N$.
We can only access the input $x\in\01^N$ by making queries to its bits.
A query is the application of the unitary transformation $O_x$ that maps
$$
O_x:\ket{i,b,z}\mapsto\ket{i,b\oplus x_i,z},
$$
where $i\in[N]$ and $b\in\01$. The $z$-part corresponds 
to the workspace, which is not affected by the query.
A $T$-query quantum algorithm has the form 
$A=U_TO_xU_{T-1}\cdots O_xU_1O_xU_0$,
where the $U_k$ are fixed unitary transformations independent of $x$. 
The final state $A\ket{0}$ depends on $x$ via the $T$ applications of $O_x$.
The output of the algorithm is determined by measuring the two rightmost
qubits of the final state. Let's say that if the rightmost bit is 1
then the algorithm claims ignorance (``don't know''), and if it is 0
then the next-to-rightmost bit is the output bit.
We refer to the survey~\cite{buhrman&wolf:dectreesurvey} 
for more details about classical and quantum query complexity.

We will use $Q_E(f)$, $Q_0(f)$, $Q_2(f)$, $NQ(f)$ to denote
the minimal query complexity of a quantum algorithm for $f$ 
in the four above modes, respectively. 
Accordingly, $Q_E(f)$ is the exact quantum query complexity of $f$, 
$Q_0(f)$ is zero-error quantum query complexity,
$Q_2(f)$ is bounded-error quantum query complexity, and
$NQ(f)$ is nondeterministic quantum query complexity.
Note that by definition we immediately have
$$
Q_2(f)\leq Q_0(f)\leq Q_E(f) \mbox{ \ and \ }
NQ(f)\leq Q_0(f)\leq Q_E(f).
$$
Our proofs will use the close connection between
quantum query complexity and polynomials~\cite{bbcmw:polynomials}.
An $N$-variate multilinear polynomial $p$ is a function of the
form $p(x)=\sum_{S\subseteq[N]}a_Sx_S$, where $a_S$ is real 
and $x_S=\prod_{i\in S}x_i$.
Its degree $deg(p)=\max\{|S|:a_S\neq 0\}$ 
is the largest degree among its monomials.
The next lemma~\cite{hoyer&wolf:disjeq,wolf:nqj} connects 
nondeterministic complexity with polynomials:

\begin{lemma}\label{lemndegpoly}
The nondeterministic quantum query complexity $NQ(f)$ of $f$
equals the minimal degree among all multilinear polynomials $p$
such that 
\begin{enumerate}
\item $p(x)\neq 0$ for all $x\in {\cal X}_1$
\item $p(x)=0$     for all $x\in {\cal X}_0$
\end{enumerate}
\end{lemma}

This lemma improves the query complexity lower bound by a factor of 2,
compared to the ``standard'' polynomial method~\cite{bbcmw:polynomials}.

The setting of computational complexity can be defined either
in terms of Turing machines or of uniform circuit families.
Here we define $\EQP$, $\ZQP$, $\BQP$, and $\NQP$ to be the classes 
of languages for which there exist polynomial-time quantum 
algorithms in the above four modes, respectively. 
We restrict attention to algebraic amplitudes for these classes.

For example, $\NQP$ (``quantum NP'') is taken to
be the class of languages $L$ for which there exists an efficient
quantum algorithm that has positive acceptance probability 
on input $x$ iff $x\in L$~\cite{adh:qcomputability}.
This class was shown to be equal to the classical counting 
class coC$_=$P~\cite{fghp:nqp,yamakami&yao:c=p}.
There is an alternative definition of quantum NP based on 
verification of quantum certificates~\cite[Chapter~14]{ksv:qc}
which we will not discuss here. We similarly define 
the classes $\EQP^A$, etc., when we have access to an oracle 
$A$ for some language, and $\EQP^S=\cup_{A\in S}\EQP^A$, etc., 
when $S$ is a set of oracles.
By definition we immediately have
$$
\EQP\subseteq\ZQP\subseteq\BQP \mbox{ \ and \ }
\EQP\subseteq\ZQP\subseteq\NQP,
$$
and these inclusions also hold relative to any oracle $A$.

\section{The problem}\label{secnonzqpf}

Let $m$ and $n$ be even numbers.
We first define the partial Boolean functions $g$ on $n$ bits and $h$ 
on $2m$ bits, and then their composition $f$ on $N=2mn$ bits.

The function $g$ is just the constant vs.~balanced problem 
of Deutsch and Jozsa~\cite{deutsch&jozsa}.
Using $w(x)$ to denote the Hamming weight of $x\in\01^n$, we define:
$$
g(x)=\left\{
\begin{array}{ll}
1, & \mbox{if }w(x)=0   \mbox{ \ \ \ (constant)}\\
0, & \mbox{if }w(x)=n/2 \mbox{ (balanced)}\\
\mbox{undefined} & \mbox{otherwise}
\end{array}
\right.
$$
It is well known that there exists an exact 1-query quantum algorithm
for this problem~\cite{deutsch&jozsa}, while any classical
deterministic or even zero-error algorithm needs $n/2+1$ queries.

The function $h$ is a zero-error sampling problem.
Let 
$$
\begin{array}{lll}
{\cal A}_1 & = & \{0^mx : x\in\01^m, m/2\leq w(x)\leq m\}\\
{\cal A}_0 & = & \{x0^m : x\in\01^m, m/2\leq w(x)\leq m\}\\[2mm]
h(x)       & = & \left\{
\begin{array}{ll}
1, & \mbox{if }x\in{\cal A}_1\\
0, & \mbox{if }x\in{\cal A}_0\\
\mbox{undefined} & \mbox{otherwise}
\end{array}\right.
\end{array}
$$
Clearly $h$ has a classical algorithm that always outputs
the correct answer and whose expected number of queries is small.
The algorithm just queries a random point in the first $m$ bits 
of its input and one in the second $m$ bits, 
and outputs where it finds a 1 (if it does so).
With probability $\geq 1/2$ it will indeed find a 1, 
so the expected number of repetitions before termination is $\leq 2$.

Let $f$ on $2mn$ bits be the partial Boolean function 
that is the composition of $g$ and $h$. 
In other words, defining the set of promise inputs by
$$
\begin{array}{lll}
{\cal X}_1 & = & \underbrace{{\cal A}_0\times\cdots\times{\cal A}_0}_{n \ \rm times}\\[7mm]
{\cal X}_0 & = & \cup\{{\cal A}_{y_1}\times\cdots\times{\cal A}_{y_n} : 
y=y_1\ldots y_n\in\01^n, w(y)=n/2\}
\end{array}
$$
we have
$$
f(x)=\left\{
\begin{array}{ll}
1, & \mbox{if }x\in{\cal X}_1 \mbox{ \ (constant)}\\
0, & \mbox{if }x\in{\cal X}_0 \mbox{ \ (balanced)}\\
\mbox{undefined} & \mbox{otherwise}
\end{array}\right.
$$
%We will use the following notation to address the $2mn$ binary variables
%in the input $x$ to $f$: $x^{(i)}_{0j}$, $i\in[n]$ and $j\in[m]$, 
%denotes the $j$th bit in the first $m$ bits of the input to 
%the $i$th copy of $h$, and $x^{(i)}_{1j}$ denotes the $j$th bit 
%in the second $m$ bits.
For later reference, we will give names to the various parts of the
$2mn$-bit input $x$:
$$
x=\overbrace{
\overbrace{\underbrace{x^{(0,1)}}_{m \rm bits}
\underbrace{x^{(1,1)}}_{m \rm bits}}^{\rm input \ for \ \it h}
\ \
\overbrace{\underbrace{x^{(0,2)}}_{m \rm bits}
\underbrace{x^{(1,2)}}_{m \rm bits}}^{\rm input \ for \ \it h}
\ \cdots\cdots\cdots\
\overbrace{\underbrace{x^{(0,n)}}_{m \rm bits}
\underbrace{x^{(1,n)}}_{m \rm bits}}^{\rm input \ for \ \it h}
}^{\rm input \ for \ \it g}
$$
In words, $f$ contains $n$ different $h$-functions, each with its
own $2m$-bit input. Here $x^{(0,i)}$ and $x^{(1,i)}$
are two $m$-bit strings that together constitute the input to 
the $i$th $h$-function.
The promise says that the $2m$-bit input $x^{(0,i)}x^{(1,i)}$
always lies in ${\cal A}_0$ or ${\cal A}_1$.
The $n$ bits $h(x^{(0,i)}x^{(1,i)})$, $i=1,\ldots,n$, coming out of the 
$n$ $h$-functions are then plugged into $g$ to give the value for $f$.
The promise says that these $n$ bits are either all 0 (constant)
or half 0 and half 1 (balanced).

Our function $f$ is just the composition of the problems $g$ and $h$,
each of which needs just a small expected number of queries.
Yet below we will show that any quantum zero-error algorithm for $f$
will need to make {\em many\/} queries.  Even stronger, also a 
{\em nondeterministic\/} quantum algorithm for $f$ requires many queries.

\section{Lower bound for quantum zero-error algorithms}

The next lemma is our main technical tool:

\begin{lemma}\label{lemzqpdegree}
Let $p$ be a $2mn$-variate multilinear polynomial such that 
\begin{enumerate}
\item $p(x)\neq 0$ for all $x\in {\cal X}_1$
\item $p(x)=0$     for all $x\in {\cal X}_0$
\end{enumerate}
Then $deg(p)\geq\min(n/2,m/2)+1$.
\end{lemma}

\begin{proof}
We use the names for the various subparts of the $2mn$-bit 
input that we introduced in Section~\ref{secnonzqpf}.
We assume without loss of generality that for every $i\in[n]$ and 
every non-zero monomial $a_Sx_S$ in $p$, the set $S$ does not 
simultaneously contain variables from $x^{(0,i)}$ and from $x^{(1,i)}$.
Since the promise on the inputs sets either $x^{(0,i)}$ or 
$x^{(1,i)}$ to $0^m$, a monomial containing variables from 
both $x^{(0,i)}$ and $x^{(1,i)}$ evaluates to 0 anyway, 
so removing it from $p$ will not affect the two properties of $p$.

Suppose, by way of contradiction, that $d=deg(p)\leq\min(n/2,m/2)$.
By the first property of the lemma, $p$ cannot be identically zero,
so it has to contain at least one monomial.
Consider a monomial $M=a_Sx_S$ in $p$ with maximal degree, so $|S|=d$. 
Consider some $i\in[n]$ such that $S$ contains variables from $x^{(1,i)}$
(and hence, by the above assumption, no variables from $x^{(0,i)}$). 
We now fix $x^{(0,i)}$ to $0^m$ and fix all non-$S$ variables 
in $x^{(1,i)}$ to 1. Since there are at most $m/2$ $S$-variables 
in total, this already sets at least $m/2$ bits in $x^{(1,i)}$ to 1.
Accordingly, we have $x^{(0,i)}x^{(1,i)}\in{\cal A}_1$ for every
setting of the $S$-variables.
This forces the $i$th $h$-function to value 1, 
without fixing the $S$-variables.  
Similarly we force the other $h$-functions whose variables intersect with $S$:
if $S$ has variables from $x^{(1,j)}$ then we force the $j$th 
$h$-function to 1, and if $S$ has variables from $x^{(0,j)}$ 
then we force it to 0. Since $|S|\leq n/2$, this forces at 
most $n/2$ of the $h$-functions. Accordingly, we can extend our
setting to the other $h$-functions (whose variables don't 
intersect with $S$ at all) to create a setting of the overall 
$2mn$-bit input that is in ${\cal X}_0$ (balanced), 
without fixing the $S$-variables.

Let $q$ be the remaining polynomial in the $d$ $S$-variables. 
No matter how we vary the $S$-variables, the overall
input to $p$ remains in ${\cal X}_0$ (balanced). Hence $q$ must 
be zero on all Boolean settings of its variables.  
It is easy to see that the only polynomial satisfying this constraint
is the one without any monomials. 
But $q$ still contains the monomial $M$, because being of degree $d$,
$M$ cannot cancel against other monomials when we fix the non-$S$ variables.  
This is a contradiction.
\end{proof}

This lemma is exactly tight. 
First, there is a polynomial with the above
properties of degree $n/2+1$. For $T$ a set of $n/2+1$ 
variables, each from a different $x^{(0,i)}$, define $q_T$ 
to be the degree-$(n/2+1)$ polynomial that is the $\AND$ 
of these variables. If $x\in{\cal X}_0$ then $q_T$ will be 0
for all $T$, and if $x\in{\cal X}_1$ then for at least one $T$ 
we have $q_T=1$. Hence summing $q_T$ over all such $T$ 
gives a polynomial $p$ of degree $n/2+1$ such that  
$p(x)=0$ for $x\in{\cal X}_0$ and $p(x)>0$ for $x\in{\cal X}_1$.

Second, there also is an appropriate polynomial of degree $m/2+1$.
Let $q_i$ be the degree-$(m/2+1)$ polynomial that is 
the $\OR$ of the first $m/2+1$ bits of $x^{(1,i)}$.
Then $q_i=1$ if $x^{(0,i)}x^{(1,i)}\in{\cal A}_1$ and 
$q_i=0$ if $x^{(0,i)}x^{(1,i)}\in{\cal A}_0$.
Defining $p$ to be the degree-$(m/2+1)$ polynomial
$n/2-\sum_{i=1}^n q_i$, we have
$p(x)=0$ for $x\in{\cal X}_0$ and $p(x)=n/2$ for $x\in{\cal X}_1$.

Combining the previous lemma with Lemma~\ref{lemndegpoly} 
gives our main theorem:

\begin{theorem}\label{thNQf}
$NQ(f)=\min(n/2,m/2)+1$.
\end{theorem}

Since nondeterministic query complexity lower bounds zero-error
complexity, we also obtain the zero-error lower bound 
$Q_0(f)\geq \min(n/2,m/2)+1$. 
The best upper bound on $Q_0(f)$ that we know, is $\min(2n,m)$
so the lower bound is tight up to small constant factors.
First, we know there is a classical zero-error algorithm that 
computes an $h$-function using an expected number of 2 queries;
we can use this to compute the first $n/2$ $h$-functions in
an expected number of $n$ queries, which suffices to compute $f$.
Terminating this algorithm after $2n$ steps gives us an algorithm 
that finds the correct output with probability $\geq 1/2$ 
(Markov's inequality), and claims ignorance otherwise.

Second, there exists an exact quantum algorithm for $f$ 
that uses $m$ queries. By querying the first $m/2$ bits 
in an $h$-input we can decide whether that $h$ takes value 0 or 1.
By copying the output and reversing the computation 
we can do this exact computation cleanly (resetting all 
workspace to 0) using $m$ queries. Putting the Deutsch-Jozsa algorithm
on top of this gives an $m$-query exact quantum algorithm for $f$.

Using a standard translation of query complexity results
to oracles, we obtain

\begin{theorem}
There exists an oracle $A$ such that 
$$
\EQP^{\rm ZPP^A}\not\subseteq \NQP^A,%\cup co\NQP^A,
$$
hence in particular
$$
\ZQP^{\rm ZQP^A}\not\subseteq \ZQP^A.
$$
\end{theorem}

\begin{proof}
For a set $A\subseteq\01^*$, we use $A^{=n}$ to denote the set of all
$n$-bit strings in $A$, and we identify this with its $2^n$-bit 
characteristic vector. We will construct a set $A$ such that, for every 
$n$ where $2^n=2m^2$ for some $m$ (i.e.~for every odd $n$), 
$A^{=n}$ is a valid input to $f$
(word of warning: the `$n$' used here is not the `$n$' used earlier,
but the `$m$' is; the input length of $f$ is now $2m^2$).  
This $A$ induces a language
$$
L=\{0^n\mid 2^n=2m^2\mbox{ for some }m\mbox{ and }f(A^{=n})=1\}.
$$
Let $M_1,M_2,\ldots$ be an enumeration of all 
oracle \NQP-machines, with increasing polynomial time bounds 
(say, $M_i$ has time bound $p_i(n)=n^i+i$). 
Such an enumeration exists because we can assume without loss 
of generality that the machines only use algebraic amplitudes
\cite{adh:qcomputability,fghp:nqp,yamakami&yao:c=p}.
At the start of our construction, $A$ is the empty set.
Going along $i=1,2,\ldots$, for each $M_i$ we will pick a specific 
input length $n_i$ and define $A^{=n_i}$ in such a way that $M_i^A$ 
will err on $0^{n_i}$, and hence it will not accept~$L$.

Consider $M_i$. Its running time is bounded by the polynomial 
$p_i(n)$ in the input length. 
Let $n_i$ be the smallest input length such that 
(1) $2^{n_i}=2m^2$ for some $m$, 
(2) $p_i(n_i)\leq m/2$, and 
(3) $n_i$ is so large that for all $j<i$ we have $p_j(n_j)<n_i$.%
\footnote{This third condition ensures that when we define $A^{=n_i}$ 
to thwart $M_i$, the behavior of earlier $M_j$s on input length $n_j$ 
won't be changed (because $M_j$ on input length $n_j$ doesn't have enough
time to query strings of length $n_i$).}
Since $M_i$ makes at most $p(n_i)<m/2+1=NQ(f)$ queries to the bits of 
$x=A^{=n_i}$, Theorem~\ref{thNQf} implies that $M_i$ cannot be 
a nondeterministic algorithm for $f$.
Hence there exists some $x\in{\cal X}_0\cup{\cal X}_1$ where $M_i$ errs:
either $x\in{\cal X}_0$ while $M_i$ has positive acceptance probability
when $A^{=n}=x$; or $x\in{\cal X}_1$ while $M_i$ has zero acceptance 
probability when $A^{=n_i}=x$. Define $A^{=n_i}$ to be that $x$. 
This ensures that $M_i^A$ does not accept $L$.

Doing this for all $M_i$ and filling the yet-undefined levels $A^{=n}$ 
by arbitrary promise-inputs to $f$, we now have a language $L$
that is accepted by none of the $M_i^A$, hence $L\not\in\NQP^A$.
On the other hand, the Deutsch-Jozsa algorithm implies 
$L\in\EQP^{\rm ZPP^A}$, so we have our separation.
\end{proof}

\section{Conclusion}

We proved that the composition of two problems that are easy
for zero-error quantum computing need not be easy itself.
This contrasts strongly with the case of classical algorithms,
and shows that our classical intuition about expected
running time does not carry over very well to quantum algorithms.
The problem in using a zero-error algorithm as a subroutine in a 
quantum algorithm seems to be that we cannot reverse the computation 
to obtain an answer without additional non-zero workspace. 
This remaining non-zero workspace then messes up later quantum interference 
in the main program. Being able to compose zero-error algorithms
is a desirable property that obviously holds in the classical world. 
Unfortunately, this property does not hold in the quantum world.

%\bibliography{qc}

\end{document}